\documentclass[a4paper]{article}
\RequirePackage[english]{babel}
\RequirePackage[latin1]{inputenc}
\RequirePackage[T1]{fontenc}
\RequirePackage{mathrsfs}
\RequirePackage{amsmath}
\RequirePackage{amssymb}
\RequirePackage{amsbsy}
\RequirePackage{bm}
\begin{document}
\title{\bf{A Discussion on Dirac Field Theory,\\ No-Go Theorems and Renormalizability}}
\author{Luca Fabbri\\
\footnotesize DIPTEM Sez. Metodi e Modelli Matematici dell'Universit\`{a} di Genova \& \\
\footnotesize INFN Sez. di Bologna and Dipartimento di Fisica dell'Universit\`{a} di Bologna}
\date{}
\maketitle
\begin{abstract}
We study Dirac field equations coupled to electrodynamics with metric and torsion fields: we discuss how special spinorial solutions are incompatible with torsion; eventually these results will be used to sketch a discussion on the problem of renormalizability of point-like particles.
\end{abstract}
\section*{Introduction}
In the present paper, we will be interested in considering the Dirac matter field, studying its dynamics in the case of its most general coupling in which electrodynamics together with metric and torsion are all present \cite{Hehl:1994ue}: for such a construction, our goal is to prove a No-Go theorem demonstrating that special solutions of the Dirac matter field are not possible whenever the Dirac spinorial matter field equations admit the presence of torsion, and torsion-spin field equations are accounted, first in Einstein, or Einstein-like, models, constructed on the Ricci scalar, or functions of the Ricci scalar, actions \cite{Capozziello:2011et}, and then in the case of conformal gravity, constructed via Weyl actions \cite{Capozziello:2009dz}; the theorem we will obtain will state that in some situation there can be no point-like field distribution, and consequently it will be brought forward to a consequent discussion on the problem of renormalizability. In order for this to be done, we will start by introducing the general formalism in terms of geometrical-kinematic bases upon which we will build the dynamics; the No-Go theorem will be proven after a preliminary result stating that in the case of stationary spherical symmetry, further generalized to the case of rotational symmetries and flat spacetimes, there can be no solution for the spinor matter field equation. The idea beneath our reasoning is that, as torsion is the spin manifested as a centrifugal potential in the spinor matter field equations, then there can be no full isotropy compatible with the particular direction selected by this centrifugal potential, and our result will simply be the rigorous proof of this intuitive fact.
\section{Dirac Field Theory}
\subsection{Establishment of Geometry and Kinematics}
In this paper, for the notations and conventions we will follow \cite{f} and we will recall here only the tools we will need: in particular, we recall that given a metric $g_{\mu\nu}$ and a connection $\Gamma^{\mu}_{\sigma\pi}$ defining a covariant derivative $D_{\mu}$ the metric-connection compatibility condition is $D_{\mu}g_{\alpha\beta}\equiv0$ and further, the Cartan torsion and Riemann curvature tensors are defined to be given by
\begin{eqnarray}
&Q^{\mu}_{\phantom{\mu}\sigma\pi}
=\Gamma^{\mu}_{\phantom{\mu}\sigma\pi}-\Gamma^{\mu}_{\phantom{\mu}\pi\sigma}
\label{torsion}\\
&G^{\mu}_{\phantom{\mu}\rho\sigma\pi}=\partial_{\sigma}\Gamma^{\mu}_{\rho\pi}
-\partial_{\pi}\Gamma^{\mu}_{\rho\sigma}
+\Gamma^{\mu}_{\lambda\sigma}\Gamma^{\lambda}_{\rho\pi}
-\Gamma^{\mu}_{\lambda\pi}\Gamma^{\lambda}_{\rho\sigma}
\label{curvature}
\end{eqnarray}
whose contractions are called the Ricci curvature tensors and scalars defined to be given by $G^{\alpha}_{\phantom{\alpha}\rho\alpha\sigma}=G_{\rho\sigma}$ and $G_{\rho\sigma}g^{\rho\sigma}=G$ fixing the convention; then the most general metric-compatible connection can be decomposed according to the form
\begin{eqnarray}
&\Gamma^{\sigma}_{\phantom{\sigma}\rho\alpha}=
\frac{1}{2}\left(Q^{\sigma}_{\phantom{\sigma}\rho\alpha}
+Q_{\rho\alpha}^{\phantom{\rho\alpha}\sigma}+Q_{\alpha\rho}^{\phantom{\alpha\rho}\sigma}\right)
+\frac{1}{2}g^{\sigma\theta}(\partial_{\rho}g_{\alpha\theta}+\partial_{\alpha}g_{\rho\theta}
-\partial_{\theta}g_{\rho\alpha})
\label{connection}
\end{eqnarray}
where the first parenthesis is in terms of torsion and called contorsion and it is a tensor while the second parenthesis is a torsionless purely metric symmetric connection $\Lambda^{\mu}_{\sigma\pi}$ defining a covariant derivative $\nabla_{\mu}$ that verifies the metric-connection compatibility condition $\nabla_{\mu}g_{\sigma\pi}\equiv0$ automatically, and where the Riemann curvature of this connection is given by the expression
\begin{eqnarray}
&0=\Lambda^{\mu}_{\phantom{\mu}\sigma\pi}-\Lambda^{\mu}_{\phantom{\mu}\pi\sigma}\\
&R^{\mu}_{\phantom{\mu}\rho\sigma\pi}=\partial_{\sigma}\Lambda^{\mu}_{\rho\pi}
-\partial_{\pi}\Lambda^{\mu}_{\rho\sigma}
+\Lambda^{\mu}_{\lambda\sigma}\Lambda^{\lambda}_{\rho\pi}
-\Lambda^{\mu}_{\lambda\pi}\Lambda^{\lambda}_{\rho\sigma}
\label{metriccurvature}
\end{eqnarray}
whose contractions are called Ricci curvature tensors and scalars and defined to be given by $R^{\alpha}_{\phantom{\alpha}\rho\alpha\sigma}=R_{\rho\sigma}$ and $R_{\rho\sigma}g^{\rho\sigma}=R$ as usual. Likewise it is possible to introduce a gauge connection $A_{\mu}$ such that its curvature is given by
\begin{eqnarray}
&F_{\mu\nu}=\partial_{\mu}A_{\nu}-\partial_{\nu}A_{\mu}
\label{strenght}
\end{eqnarray}
which is irreducible and containing no torsion at all.

In order to introduce matter fields as well, we have to translate everything in the tetrad formalism: so by taking into account a pair of bases of vectors called vierbeins $\xi^{a}_{\sigma}$ and $\xi_{a}^{\sigma}$ dual of one another $\xi^{a}_{\mu}\xi_{a}^{\rho}=\delta^{\rho}_{\mu}$ and $\xi^{a}_{\mu}\xi_{r}^{\mu}=\delta_{r}^{a}$ it is always possible to take them orthonormal $\xi^{a}_{\sigma} \xi^{q}_{\rho} g^{\sigma\rho}=\eta^{aq}$ or $\xi_{a}^{\sigma} \xi_{q}^{\rho} g_{\sigma\rho}=\eta_{aq}$ where we have that $\eta_{aq}$ and $\eta^{aq}$ are unitary diagonal matrices called Minkowskian matrices whereas the spin-connection $\varsigma^{ij}_{\phantom{ij}\mu}$ defines a covariant derivative $D_{\mu}$ for which conditions $D_{\mu}\xi_{a}^{\sigma}\equiv0$ and $D_{\mu}\eta_{ab}\equiv0$ are imposed; these two conditions are equivalent respectively to the two conditions given by the expression
\begin{eqnarray}
&\varsigma^{ij}_{\phantom{ij}\alpha}=
\eta^{jp}e^{\rho}_{p}e^{i}_{\sigma}\Gamma^{\sigma}_{\rho\alpha}
+\eta^{jp}e^{i}_{\sigma}\partial_{\alpha}e^{\sigma}_{p}
\label{spin-connection}
\end{eqnarray} 
and the property $\varsigma^{ip}_{\phantom{ip}\alpha}=-\varsigma^{pi}_{\phantom{pi}\alpha}$, showing that the antisymmetric spin-connection can be written in terms of the connection and therefore decomposed in terms of torsional and torsionless spin-connection $\omega^{ij}_{\phantom{ij}\mu}$ defining an analogous covariant derivative $\nabla_{\mu}$ that verifies $\nabla_{\mu} \xi^{a}_{\sigma}\equiv0$ and $\nabla_{\mu} \eta_{ab}\equiv0$ automatically, and where
\begin{eqnarray}
&-Q^{a}_{\phantom{a}\alpha\rho}
=\partial_{\alpha}\xi^{a}_{\rho}-\partial_{\rho}\xi^{a}_{\alpha}
+\xi^{k}_{\rho}\varsigma^{a}_{\phantom{a}k\alpha}
-\xi^{k}_{\alpha}\varsigma^{a}_{\phantom{a}k\rho}\\
&G^{a}_{\phantom{a}b\sigma\pi}=\partial_{\sigma}\varsigma^{a}_{\phantom{a}b\pi}
-\partial_{\pi}\varsigma^{a}_{\phantom{a}b\sigma}
+\varsigma^{a}_{\phantom{a}k\sigma}\varsigma^{k}_{\phantom{k}b\pi}
-\varsigma^{a}_{\phantom{a}k\pi}\varsigma^{k}_{\phantom{k}b\sigma}
\end{eqnarray}
is the expression of torsion and curvature in this formalism. In this form it is actually straightforward to see that in this formalism, torsion and curvature are the strength respectively of vierbein and spin-connection if these are thought as potentials, as the tensor $F_{\mu\nu}$ is the strength of the $A_{\mu}$ potential, showing the analogies between the underlying theories, where the vierbein $\xi^{i}_{\alpha}$ and spin-connection $\varsigma^{k}_{\phantom{k}b\sigma}$ are the potentials of respectively spacetime translations and rotations and the vector $A_{\mu}$ is the potential of phase transformations, when spacetime rototraslations and phase tranformations are gauged as in Poincar\'{e} and Maxwell gauge theories of gravity and electrodynamics, mentioned above.

To write the above formalism in spinorial representation, we have to introduce the set of complex matrices $\gamma_{a}$ defined as to belong to the well known Clifford algebra $\{\gamma_{i},\gamma_{j}\}=2\mathbb{I}\eta_{ij}$ so that we can build $[\gamma_{i},\gamma_{j}]=4\sigma_{ij}$ being such that $\{\gamma_{i},\sigma_{jk}\}= i\varepsilon_{ijkq}\gamma\gamma^{q}$ and $[\gamma_{i},\sigma_{jk}]=\eta_{ij}\gamma_{k}-\eta_{ik}\gamma_{j}$ where the $\sigma_{jk}$ matrices are the generators of the spinorial transformation $S$ as known: a spinor will be defined as what transforms according to $S$ and the passage between spinors and conjugate spinors $\psi$ and $\overline{\psi}$ is given in terms of $\gamma_{0}$ as $\overline{\psi}\equiv \psi^{\dagger}\gamma_{0}$ while the introduction of the spinorial connection $\Sigma_{\mu}$ is what allows the definition of the spinorial derivative $D_{\mu}\psi\equiv \partial_{\mu}\psi +\Sigma_{\mu}\psi$ yielding the conditions $D_{\mu}\gamma_{a}\equiv0$ by construction; these conditions allow us to employ the spin-connection and the gauge field to write the most general form of the spinorial connection as
\begin{eqnarray}
&\Sigma_{\alpha}=\frac{1}{2}\varsigma^{ij}_{\phantom{ij}\alpha}\sigma_{ij}+iqA_{\alpha}\mathbb{I}
\label{spinorial-connection}
\end{eqnarray} 
for any parameter $q$ completely general, and when torsion is separated apart the torsionless quantities are indicated by the spinorial connection $\Omega_{\mu}$ defining a spinorial derivative $\nabla_{\mu}$ for which $\nabla_{\mu} \gamma_{a}\equiv0$ hold, and with curvature
\begin{eqnarray}
&\Gamma_{\sigma\pi}=\partial_{\sigma}\varsigma_{\pi}-\partial_{\pi}\varsigma_{\sigma}
+\varsigma_{\sigma}\varsigma_{\pi}-\varsigma_{\pi}\varsigma_{\sigma}
\end{eqnarray}
containing the spacetime curvature with torsion and the Maxwell field \cite{f}.
\subsection{Setting the Dynamics}
After this introduction of the geometrical background, we may write the least-order derivative dynamics given by the usual Dirac matter lagrangian
\begin{eqnarray}
&\mathcal{L}=G-\frac{1}{4}F^{2}
+\frac{i}{2}\left(\overline{\psi}\gamma^{\mu}D_{\mu}\psi
-D_{\mu}\overline{\psi}\gamma^{\mu}\psi\right)-m\overline{\psi}\psi
\end{eqnarray}
where $m$ is the mass of the matter field: its variation gives torsion-spin coupling
\begin{eqnarray}
&Q^{\rho\mu\nu}
=-\frac{i}{4}\overline{\psi}\{\gamma^{\rho},\sigma^{\mu\nu}\}\psi
\end{eqnarray}
and the curvature-energy coupling field equations
\begin{eqnarray}
&G^{\mu}_{\phantom{\mu}\nu}-\frac{1}{2}\delta^{\mu}_{\nu}G
-\frac{1}{8}\delta^{\mu}_{\nu}F^{2}+\frac{1}{2}F^{\rho\mu}F_{\rho\nu}
=\frac{i}{4}\left(\overline{\psi}\gamma^{\mu}D_{\nu}\psi
-D_{\nu}\overline{\psi}\gamma^{\mu}\psi\right)
\end{eqnarray}
and also the gauge-current coupling field equations
\begin{eqnarray}
&D_{\sigma}F^{\sigma\rho}+\frac{1}{2}F_{\mu\nu}Q^{\mu\nu\rho}
=q\left(\overline{\psi}\gamma^{\rho}\psi\right)
\end{eqnarray}
together with the matter field equation
\begin{eqnarray}
&i\gamma^{\mu}D_{\mu}\psi-m\psi=0
\end{eqnarray}
in terms of the charge $q$ and mass $m$ of the matter field; in the form given above we have written all curvatures and derivatives in terms of the full torsional connection, but after torsion is separated away, we may employ the torsion-spin coupling algebraic relationship given by the above expression
\begin{eqnarray}
&Q^{\rho\mu\nu}
=-\frac{i}{4}\overline{\psi}\{\gamma^{\rho},\sigma^{\mu\nu}\}\psi
\end{eqnarray}
to write torsion in terms of the spin in all other field equations, so that we are left with a torsionless system of field equations for the curvature-energy coupling
\begin{eqnarray}
\nonumber
&R_{\mu\nu}-\frac{1}{8}g_{\mu\nu}F^{2}+\frac{1}{2}g^{\eta\rho}F_{\mu\eta}F_{\nu\rho}
=-\frac{1}{4}m\overline{\psi}\psi g_{\mu\nu}+\\
&+\frac{i}{8}\left(\overline{\psi}\gamma_{\mu}\nabla_{\nu}\psi
+\overline{\psi}\gamma_{\nu}\nabla_{\mu}\psi
-\nabla_{\nu}\overline{\psi}\gamma_{\mu}\psi
-\nabla_{\mu}\overline{\psi}\gamma_{\nu}\psi\right)
\end{eqnarray}
for the gauge-current coupling
\begin{eqnarray}
&\nabla_{\sigma}F^{\sigma\rho}=q\left(\overline{\psi}\gamma^{\rho}\psi\right)
\end{eqnarray}
and for the matter field
\begin{eqnarray}
\nonumber
&i\gamma^{\mu}\nabla_{\mu}\psi
+\frac{3}{16}\left(\overline{\psi}\gamma\gamma_{\mu}\psi\right)\gamma\gamma^{\mu}\psi-m\psi
\equiv i\gamma^{\mu}\nabla_{\mu}\psi
-\frac{3}{16}\left(\overline{\psi}\gamma_{\mu}\psi\right)\gamma^{\mu}\psi-m\psi\equiv\\
&\equiv i\gamma^{\mu}\nabla_{\mu}\psi
-\frac{3}{16}\left[i\left(i\overline{\psi}\gamma\psi\right)\gamma
+\left(\overline{\psi}\psi\right)\mathbb{I}\right]\psi-m\psi=0
\end{eqnarray}
in which for the curvature-energy and gauge-current coupling the field equations are formally equal to the field equations we have in the torsionless case, whereas for the matter field the field equations are equivalent to the field equations in the torsionless case with additional potentials given by $i\overline{\psi}\gamma\psi$ and $\overline{\psi}\psi$ as self-interactions of the fields with themselves. Now beside all contributions that clearly come from the very presence of an additional field equation for torsion, we also want torsion to maintain relevant contributions within the Dirac field equations and therefore we are going to insist on the fact that of the two conditions given by $i\overline{\psi}\gamma\psi\neq0$ or $\overline{\psi}\psi\neq0$ one is always verified for the system.

This construction is based on an action whose gravitational part is the Einstein action given in terms of the Ricci scalar \cite{f}: we will see that this action can be generalized to generic functions of the Ricci scalar \cite{Fabbri:2010pk}; we will also study conformally invariant actions given in terms of the Weyl curvature \cite{Fabbri:2011ha}.
\section{No-Go Theorems}
\subsection{Stationary Spherical Symmetry}
For the system of Dirac field equations, solutions are expected to display some localization; and accordingly stable compact solutions should find place in stationary spherically symmetric spacetimes: in the past there have been discussions on this physical situation, and in such background symmetries Dirac matter fields have been considered in \cite{c-c-d-n,m-r,f-s-y/1}, a more comprehensive case including gauge potentials has been studied in \cite{l,f-s-y/3}, while the extension accounting for gravitational effects has been studied in \cite{f-s-y/5} and the most complete situation in which also torsion is considered is our goal in the following of this paper.

Therefore, we are going to consider Dirac fields with electrodynamics and gravitation with torsion in this background, that is we will take into account the dynamics of the Dirac matter field accounting for the full coupling in a background with stationary spherical symmetry: in the frame at rest with respect to the origin of coordinates $(t,r,\theta,\varphi)$ the metric is given in terms of two functions of the radial coordinate $A(r)$ and $B(r)$ in the following form
\begin{eqnarray}
&g_{tt}=A^{2}\ \ \ \ \ \ g_{rr}=-B^{2}\ \ \ \ \ \ 
g_{\theta\theta}=-r^{2}\ \ \ \ \ \ g_{\varphi\varphi}=-r^{2}(\sin{\theta})^{2}
\end{eqnarray}
while the tetrads are given by
\begin{eqnarray}
&e^{0}_{t}=A\ \ \ \ \ \ \ \ e^{1}_{r}=B\ \ \ \ \ \ \ \ 
e^{2}_{\theta}=r\ \ \ \ \ \ \ \ e^{3}_{\varphi}=r\sin{\theta}
\end{eqnarray}
with gamma matrices in chiral representation; the symmetric connection is
\begin{eqnarray}
\nonumber
&\Lambda^{t}_{tr}=\frac{A'}{A}\ \ \ \ \ \ \ \ \ \ \ \ 
\Lambda^{r}_{tt}=\frac{AA'}{B^{2}}\ \ \ \ \ \ \ \ 
\Lambda^{r}_{rr}=\frac{B'}{B}\\
\nonumber
&\Lambda^{r}_{\theta\theta}=-\frac{r}{B^{2}}\ \ \ \ \ \ \ \ \ \ \ \ 
\Lambda^{r}_{\varphi\varphi}=-\frac{r}{B^{2}}(\sin{\theta})^{2}\\
\nonumber
&\Lambda^{\theta}_{\theta r}=\frac{1}{r}\ \ \ \ \ \ \ \ 
\Lambda^{\theta}_{\varphi\varphi}=-\cot{\theta}(\sin{\theta})^{2}\\
&\Lambda^{\varphi}_{\varphi r}=\frac{1}{r}\ \ \ \ \ \ \ \ \ \ 
\Lambda^{\varphi}_{\varphi\theta}=\cot{\theta}
\end{eqnarray}
whereas the spin connection is given by
\begin{eqnarray}
\nonumber
&\omega^{0}_{\phantom{0}1t}=\frac{A'}{B}\ \ \ \ 
\omega^{1}_{\phantom{1}2\theta}=-\frac{1}{B}\ \ \ \ 
\omega^{1}_{\phantom{1}3\varphi}=-\frac{1}{B}\sin{\theta}\\  &\omega^{2}_{\phantom{2}3\varphi}=-\cos{\theta}
\end{eqnarray}
and for which the spinorial connection is
\begin{eqnarray}
\nonumber
&\Omega_{t}=\frac{A'}{2B}\gamma_{1}\gamma_{0}\ \ \ \Omega_{r}=0\ \ \ 
\Omega_{\theta}=\frac{1}{2B}\gamma_{1}\gamma_{2}\\
&\Omega_{\varphi}=\left(\frac{1}{2B}\gamma_{1}
+\frac{1}{2}\cot{\theta}\gamma_{2}\right)\sin{\theta}\gamma_{3}
\end{eqnarray}
as it may be checked in common textbooks. The electrodynamic potential producing electrostatic radial configurations is $A_{t}=\phi$ with $\phi=\phi(r)$ as usual.

The Dirac matter fields with electrodynamics and gravitation with torsion can be written in this background: then we have field equations given by
\begin{eqnarray}
\nonumber
&i\left(m+\frac{3}{16}\overline{\psi}\psi\right)\left(\sqrt{Ar^{2}\sin{\theta}}\psi\right)
-\gamma\left(\frac{3}{16}i\overline{\psi}\gamma\psi\right)
\left(\sqrt{Ar^{2}\sin{\theta}}\psi\right)+\\
\nonumber
&+\gamma_{0}\frac{1}{A}\left(iq\phi+\frac{\partial}{\partial t}\right)
\left(\sqrt{Ar^{2}\sin{\theta}}\psi\right)
-\gamma_{1}\frac{1}{B}\frac{\partial}{\partial r}\left(\sqrt{Ar^{2}\sin{\theta}}\psi\right)-\\
&-\gamma_{2}\frac{1}{r}\frac{\partial}{\partial\theta}\left(\sqrt{Ar^{2}\sin{\theta}}\psi\right)
-\gamma_{3}\frac{1}{r\sin{\theta}}\frac{\partial}{\partial\varphi}\left(\sqrt{Ar^{2}\sin{\theta}}\psi\right)=0
\label{fieldequation}
\end{eqnarray}
and these are the field equations we are going to employ in the following study.

To this extent, we work in the chiral representation, that is the representation in which the reducibility of the field $\psi$ in left/right-handed projections becomes manifest, and in the case of spherical symmetry they correspond to spherical waves that may either be incoming to or outgoing from the origin, although the fact that these two waves are equivalent up to a time reversal allows us to pick only one of them, while still being sure that the results we obtain will hold for the other as well: choosing the outgoing wave, the most general spinor is
\begin{eqnarray}
&\psi=\left(\begin{tabular}{c}
$\rho e^{i\frac{\alpha}{2}}$\\
$\rho e^{i\frac{\alpha}{2}}$\\
$\eta e^{i\frac{\beta}{2}}$\\
$\eta e^{i\frac{\beta}{2}}$
\end{tabular}\right)
\label{fieldsolution}
\end{eqnarray}
as this is the most general form for which the corresponding bilinear fields
\begin{eqnarray}
&\overline{\psi}\gamma^{0}\psi=\overline{\psi}\gamma^{1}\gamma\psi
=2\left(\eta^{2}+\rho^{2}\right)\\
&\overline{\psi}\gamma^{1}\psi=\overline{\psi}\gamma^{0}\gamma\psi
=2\left(\eta^{2}-\rho^{2}\right)\\
&i\overline{\psi}\gamma\psi=4\eta\rho\sin{\left(\frac{\alpha-\beta}{2}\right)}\\
&\overline{\psi}\psi=4\eta\rho\cos{\left(\frac{\alpha-\beta}{2}\right)} 
\end{eqnarray}
are the most general to be compatible with the rotational invariance, and from these it is also possible to construct additional scalar and pseudo-scalar bilinears
\begin{eqnarray}
&\overline{\psi}\gamma^{a}\psi\overline{\psi}\gamma_{a}\psi
=-\overline{\psi}\gamma^{a}\gamma\psi\overline{\psi}\gamma_{a}\gamma\psi
=\left(\overline{\psi}\psi\right)^{2}+\left(i\overline{\psi}\gamma\psi\right)^{2}
=16\eta^{2}\rho^{2}
\label{scalar}
\end{eqnarray}
which depend on the radial coordinate only, with the product $\eta\rho$ and the difference $\alpha-\beta$ depending on the radial coordinate solely. These constraints come from having imposed that not only the vector $\overline{\psi}\gamma^{a}\psi$ related to the momentum but also the pseudo-vector $\overline{\psi}\gamma_{a}\gamma\psi$ related to the spin must be both restricted to the spherically symmetric case, while conditions $i\overline{\psi}\gamma\psi\neq0$ or $\overline{\psi}\psi\neq0$ imply we have to require both $\eta\neq0$ and $\rho\neq0$ in the most general case we will discuss.

By plugging into the field equation (\ref{fieldequation}) the field given by (\ref{fieldsolution}) we get
\begin{eqnarray}
\nonumber
&\left(m+\frac{3}{4}\eta\rho\cos{\left(\frac{\alpha-\beta}{2}\right)}\right)
\left(\eta e^{i\frac{\beta}{2}}\right)
-i\left(\frac{3}{4}\eta\rho\sin{\left(\frac{\alpha-\beta}{2}\right)}\right)
\left(-\eta e^{i\frac{\beta}{2}}\right)-\\
\nonumber
&-\frac{1}{\sqrt{Ar^{2}}}\left(\frac{i}{A}\left(iq\phi+\frac{\partial}{\partial t}\right)
\left(-\sqrt{Ar^{2}}\rho e^{i\frac{\alpha}{2}}\right)
+\frac{i}{B}\frac{\partial}{\partial r}\left(\sqrt{Ar^{2}}\rho e^{i\frac{\alpha}{2}}\right)\right)-\\
&-\frac{1}{r}\left[\frac{1}{\sqrt{\sin{\theta}}}\left(\frac{\partial}{\partial\theta}\left(\sqrt{\sin{\theta}}\rho e^{i\frac{\alpha}{2}}\right)
+\frac{i}{\sin{\theta}}\frac{\partial}{\partial\varphi}\left(\sqrt{\sin{\theta}}\rho e^{i\frac{\alpha}{2}}\right)\right)\right]=0
\label{1}\\
\nonumber
&\left(m+\frac{3}{4}\eta\rho\cos{\left(\frac{\alpha-\beta}{2}\right)}\right)
\left(-\eta e^{i\frac{\beta}{2}}\right)
-i\left(\frac{3}{4}\eta\rho\sin{\left(\frac{\alpha-\beta}{2}\right)}\right)
\left(\eta e^{i\frac{\beta}{2}}\right)-\\
\nonumber
&-\frac{1}{\sqrt{Ar^{2}}}\left(\frac{i}{A}\left(iq\phi+\frac{\partial}{\partial t}\right)
\left(\sqrt{Ar^{2}}\rho e^{i\frac{\alpha}{2}}\right)
+\frac{i}{B}\frac{\partial}{\partial r}\left(-\sqrt{Ar^{2}}\rho e^{i\frac{\alpha}{2}}\right)\right)-\\
&-\frac{1}{r}\left[\frac{1}{\sqrt{\sin{\theta}}}\left(\frac{\partial}{\partial\theta}\left(\sqrt{\sin{\theta}}\rho e^{i\frac{\alpha}{2}}\right)
+\frac{i}{\sin{\theta}}\frac{\partial}{\partial\varphi}\left(\sqrt{\sin{\theta}}\rho e^{i\frac{\alpha}{2}}\right)\right)\right]=0
\label{2}\\
\nonumber
&\left(m+\frac{3}{4}\eta\rho\cos{\left(\frac{\alpha-\beta}{2}\right)}\right)
\left(-\rho e^{i\frac{\alpha}{2}}\right)
-i\left(\frac{3}{4}\eta\rho\sin{\left(\frac{\alpha-\beta}{2}\right)}\right)
\left(-\rho e^{i\frac{\alpha}{2}}\right)-\\
\nonumber
&-\frac{1}{\sqrt{Ar^{2}}}\left(\frac{i}{A}\left(iq\phi+\frac{\partial}{\partial t}\right)
\left(\sqrt{Ar^{2}}\eta e^{i\frac{\beta}{2}}\right)
+\frac{i}{B}\frac{\partial}{\partial r}\left(\sqrt{Ar^{2}}\eta e^{i\frac{\beta}{2}}\right)\right)-\\
&-\frac{1}{r}\left[\frac{1}{\sqrt{\sin{\theta}}}\left(\frac{\partial}{\partial\theta}\left(\sqrt{\sin{\theta}}\eta e^{i\frac{\beta}{2}}\right)
+\frac{i}{\sin{\theta}}\frac{\partial}{\partial\varphi}\left(\sqrt{\sin{\theta}}\eta e^{i\frac{\beta}{2}}\right)\right)\right]=0
\label{3}\\
\nonumber
&\left(m+\frac{3}{4}\eta\rho\cos{\left(\frac{\alpha-\beta}{2}\right)}\right)
\left(\rho e^{i\frac{\alpha}{2}}\right)
-i\left(\frac{3}{4}\eta\rho\sin{\left(\frac{\alpha-\beta}{2}\right)}\right)
\left(\rho e^{i\frac{\alpha}{2}}\right)-\\
\nonumber
&-\frac{1}{\sqrt{Ar^{2}}}\left(\frac{i}{A}\left(iq\phi+\frac{\partial}{\partial t}\right)
\left(-\sqrt{Ar^{2}}\eta e^{i\frac{\beta}{2}}\right)
+\frac{i}{B}\frac{\partial}{\partial r}\left(-\sqrt{Ar^{2}}\eta e^{i\frac{\beta}{2}}\right)\right)-\\
&-\frac{1}{r}\left[\frac{1}{\sqrt{\sin{\theta}}}\left(\frac{\partial}{\partial\theta}\left(\sqrt{\sin{\theta}}\eta e^{i\frac{\beta}{2}}\right)
+\frac{i}{\sin{\theta}}\frac{\partial}{\partial\varphi}\left(\sqrt{\sin{\theta}}\eta e^{i\frac{\beta}{2}}\right)\right)\right]=0
\label{4}
\end{eqnarray}
as it is easy to check by performing a direct substitution.

It is now straightforward to see that by adding equation (\ref{1}) to (\ref{2}) and equation (\ref{3}) to (\ref{4}) one gets the coupled system of field equations
\begin{eqnarray}
&\cot{\theta}+\frac{1}{\rho^{2}}\frac{\partial \rho^{2}}{\partial\theta}
-\frac{1}{\sin{\theta}}\frac{\partial\alpha}{\partial\varphi}=0\ \ \ \ \ \ \ \ 
&\frac{\partial\alpha}{\partial\theta}
+\frac{1}{\rho^{2}}\frac{1}{\sin{\theta}}\frac{\partial \rho^{2}}{\partial\varphi}=0\\
&\cot{\theta}+\frac{1}{\eta^{2}}\frac{\partial \eta^{2}}{\partial\theta}
-\frac{1}{\sin{\theta}}\frac{\partial\beta}{\partial\varphi}=0\ \ \ \ \ \ \ \ 
&\frac{\partial\beta}{\partial\theta}
+\frac{1}{\eta^{2}}\frac{1}{\sin{\theta}}\frac{\partial\eta^{2}}{\partial\varphi}=0
\end{eqnarray}
for the angular fields, which can be expanded and recombined to give
\begin{eqnarray}
&\frac{\partial}{\partial\theta}\left(\alpha+\beta\right)=0\ \ \ \ \ \ \ \ 
&2\cos{\theta}-\frac{\partial}{\partial\varphi}\left(\alpha+\beta\right)=0
\end{eqnarray}
since from (\ref{scalar}) we know that $\eta\rho$ and $\alpha-\beta$ do not depend on any angle, so that finally we see that the derivative of the first with respect to the azimuthal angle and the derivative of the second with respect to the elevation angle combine to give the constrain $\sin{\theta}=0$ which can not be valid if the elevation angle is to have a general value: and consequently we have obtained a contradiction.

So stationary spherically symmetric spacetimes and torsionally-interacting spinorial matter fields cannot be reciprocally compatible.

It is possible to extend these results to the case in which static spherically symmetric spacetimes are not only isotropic with respect to a single point but to every point and therefore also homogeneous: in the frame of the comoving coordinate system $(t,r,\theta,\varphi)$ the metric is given in terms of the spatial curvature $k$ and one function of time $A(t)$ in the following form
\begin{eqnarray}
&g_{tt}=1\ \ \ \ g_{rr}=-\frac{A^{2}}{\left(1-kr^{2}\right)}\ \ \ \ 
g_{\theta\theta}=-A^{2}r^{2}\ \ \ \ g_{\varphi\varphi}=-A^{2}r^{2}(\sin{\theta})^{2}
\end{eqnarray}
and we do not need to calculate the tetrads; the symmetric connection is
\begin{eqnarray}
\nonumber
&\Lambda^{t}_{rr}=\frac{A\dot{A}}{\left(1-kr^{2}\right)}\ \ \ \ \ \ 
\Lambda^{t}_{\theta\theta}=A\dot{A}r^{2}\ \ \ \ \ \ 
\Lambda^{t}_{\varphi\varphi}=A\dot{A}r^{2}(\sin{\theta})^{2}\\
\nonumber
&\Lambda^{r}_{rt}=\frac{\dot{A}}{A}\ 
\Lambda^{r}_{rr}=\frac{kr}{\left(1-kr^{2}\right)}\ 
\Lambda^{r}_{\theta\theta}=\left(kr^{3}-r\right)\ 
\Lambda^{r}_{\varphi\varphi}=\left(kr^{3}-r\right)(\sin{\theta})^{2}\\
\nonumber
&\Lambda^{\theta}_{\theta t}=\frac{\dot{A}}{A}\ \ \ \ \ \ \ \ 
\Lambda^{\theta}_{\theta r}=\frac{1}{r}\ \ \ \ \ \ \ \ 
\Lambda^{\theta}_{\varphi\varphi}=-\cot{\theta}(\sin{\theta})^{2}\\
&\Lambda^{\varphi}_{\varphi t}=\frac{\dot{A}}{A}\ \ \ \ \ \ 
\Lambda^{\varphi}_{\varphi r}=\frac{1}{r}\ \ \ \ \ \ 
\Lambda^{\varphi}_{\varphi\theta}=\cot{\theta}
\end{eqnarray}
and we do not need to calculate the spin-connection nor the Fock-Ivanenko coefficients as the result is obtained without employing dynamical field equations.

In this case in fact, we are not even going to use the field equations, but only the algebraic structure of the spinor field $\psi$ to see that for any representation the bilinear fields $\overline{\psi}\gamma^{\mu}\psi$ and $\overline{\psi}\gamma_{\mu}\gamma\psi$ may only possess the temporal components in order to be compatible with the roto-translational invariance: then the spinorial identity $\overline{\psi}\gamma^{\mu}\psi\overline{\psi}\gamma_{\mu}\psi +\overline{\psi}\gamma^{\mu}\gamma\psi\overline{\psi}\gamma_{\mu}\gamma\psi\equiv0$ immediately provides the relationship given by $|\overline{\psi}\gamma^{0}\psi|^{2} 
+|\overline{\psi}\gamma^{0}\gamma\psi|^{2}\equiv\overline{\psi}\gamma^{\mu}\psi \overline{\psi}\gamma_{\mu}\psi+\overline{\psi}\gamma^{\mu}\gamma\psi \overline{\psi}\gamma_{\mu}\gamma\psi \equiv0$ implying in particular that $\psi^{\dagger}\psi\equiv\overline{\psi}\gamma^{0}\psi=0$ but because the quantity $\psi^{\dagger}\psi=0$ is strictly positive defined we have that the solution $\psi=0$ alone is ultimately possible.

So also roto-translationally symmetric spacetimes and torsionally-interacting spinorial matter fields cannot be compatible once again.

It is even easier to see how we can extend these results to the case where isotropic-homogeneous spacetimes are Minkowskian spacetimes: in this case in fact, no dynamical consideration nor algebraic argument are needed, since the very presence of the two vectors $\overline{\psi}\gamma^{\mu}\psi$ and $\overline{\psi}\gamma^{\mu}\gamma\psi$ is enough to see that the system cannot be compatible with the Poincar\'{e} symmetry of the spacetime.

So eventually even flat spacetimes and torsionally-interacting spinorial matter fields cannot be compatible, as it is expected.

In fact, we have to remark that the results here presented have been obtained by exploiting only the rotational invariance, and they can only be empowered when also the translational invariance is implemented; this means that the incompatibility with the Lorentz symmetry is straightforwardly extended to the incompatibility with the Poincar\'{e} symmetry: this result can be easily interpreted by thinking that the presence of a spinning particle in a region of the spacetime clearly makes the spinning axis and that region privileged with respect to any other direction and position. However, these results must not be misinterpreted by thinking that physics is not compatible with the rototraslational symmetry, and as mentioned above physics is not only compatible with rototraslations but the theory of gravity with curvature and torsion is obtained precisely when the Poincar\'{e} symmetry is gauged; but the fact that the theory of gravity with curvature and torsion has field equations respecting Poincar\'{e} symmetry does not imply that all of its solutions must. This is not surprising, and the fact that the symmetry of the equations is not necessarily maintained by some solutions is a widely known fact, and examples are found not only in gravity but also for the physics of the weak interactions and in condensed matter and superconductors, all examples of symmetries spontaneously broken.

As we have already mentioned, although these results have been obtained in the case of torsional-interactions for Dirac field equations of the least-order derivative, we may now proceed to study higher-order derivative dynamics, for instance by postulating that the gravitational action is not the simplest Ricci scalar but a general function $F$ of the Ricci scalar \cite{Fabbri:2010pk}: in this reference it has been shown that the matter field equations are given by the following
\begin{eqnarray}
&i\gamma^{\mu}\nabla_{\mu}\psi
-\frac{3}{16F'}\left[i\left(i\overline{\psi}\gamma\psi\right)\gamma
+\left(\overline{\psi}\psi\right)\mathbb{I}\right]\psi-m\psi=0
\end{eqnarray}
with $m$ the mass of the matter field, equivalent to the field equations in the torsionless case obtained before with non-linear potentials given by self-interactions of the fields with themselves as $i\overline{\psi}\gamma\psi$ and $\overline{\psi}\psi$ with running coupling given by the energy-dependent scale factor $F'$ at denominator. So we are going to assume that either $i\overline{\psi}\gamma\psi\neq0$ or $\overline{\psi}\psi\neq0$ and $F'\neq1$ to ensure that in the Dirac field equations the torsion tensor in this extended gravity has relevant contributions.

With these assumptions, the Ricci scalar is a function of either the radial or the temporal coordinate, and any function of the Ricci scalar will itself be a function of either the radial or temporal coordinate, and then only the radial-temporal field equations will be changed, but the angular field equations will remain unmodified: consequently, since all result obtained here were based on the angular field equations they will also be obtained in this extended case.

All these results can be collected together into a single scheme, as a sort of no-go theorem stating that Dirac matter field equations in the most general coupling are not compatible with any rotationally invariant background.

As a particular isotropic solution is the Dirac delta distribution $\delta(r)$ we have the no-go corollary for which Dirac matter field equations in the most general coupling have no Dirac delta distributions as solutions.

Now, all these results have been obtained in the case of Dirac matter field equations in the framework in which the gravitational action was a function of the Ricci scalar alone, and further studies may be done in the case of Dirac massless field equations in a framework of a gravitational action that is conformally invariant, as discussed in \cite{Fabbri:2011ha}: in this reference it has been shown that the Dirac massless field equations are written in the usual form but, since the algebraic torsion-spin coupling is lost, torsion cannot be replaced with the spin density of the spinor field, and the spinor field self-interacting potential is absent.

As the spinor field self-interacting potential is what forbids isotropic spacetimes arising from the rotational invariance of the background to be permitted, the fact that such a potential is not present allows massless fields to have singular delta distributions as possible solutions, and the no-go theorem cannot be extended to the case of conformal fields straightforwardly.
\section{Renormalizability}
So far we have proven a no-go theorem stating that for the Dirac least-order derivative field equations in the most general coupling in which not only electrodynamics but also gravity with torsion is accounted within rotationally invariant background there is no solution respecting those symmetries, and that this result can be extended from rotationally invariant to rototraslationally invariant or even thoroughly flat backgrounds; it can also be extended to backgrounds in which gravity with torsion is dynamically defined in terms of a gravitational action that is not the Ricci scalar but a function of the Ricci scalar: in all these cases there is a no-go corollary stating that the Dirac delta distribution cannot be a solution. We have also discussed how the Dirac massless field equations coupled to electrodynamics and gravity with torsion in conformal models cannot extend the previous results and therefore solutions compatible with the rotational symmetry may exist: in this case there is not a no-go corollary stating that the Dirac delta distribution is not a solution as well. This result is important for the issue of non-renormalizability of fermions as point-like particles.

The non-renormalizability of fermions when they are considered as point-like particles, and their fermionic interactions when considered as zero-range forces, is due to the fact that it is impossible to control the ultraviolet divergences occurring at high energies for the fermion because of the absence of a cut-off for point-like objects: for the Dirac massive field least-order derivative equations in backgrounds in which the gravitational action is function of the Ricci scalar we have seen that because fermion fields have fermionic interactions they cannot be point-like, so that the cut-off for the field is provided by the intrinsic size of the spinning extended solution \cite{p,i,lo}, so that between finite-size fields even spin contact forces have finite-range \cite{b,e}; for the Dirac massless field equations in conformal models, because of the absence of fermionic interactions point-like solutions might still be admitted, although in this case renormalizability is ensured by the scale invariance itself \cite{s}. Inasmuch as our discussion is focused on either Einsteinian-like or conformal gravity, the dilemma about non-renormalizability may either be avoided or circumvented, respectively.

If non-renormalizability comes from having fermions as described by point-like particles, this form of non-renormalizability is no issue. However other types of non-renormalizability would have to be studied independently.

Nevertheless, it may be thinkable that also in those cases the method used in this paper may be of any help.
\section*{Conclusion}
In the present paper, we have studied the Dirac matter field equations in the most general coupling in which torsion is considered beside the metric and together with the electrodynamic field: we have shown that in Einstein gravity, torsion is not compatible with rotationally-invariant spacetimes, whether they are isotropic, isotropic-homogeneous, flat backgrounds; if generalizations of the Einstein gravity are considered to be the Einstein-like gravity, all results on torsion incompatibility with such backgrounds are unchanged: in particular material fields described by delta distributions representing singular solutions cannot be admitted; on the other hand, in conformal gravity, torsion may be compatible with rotationally-invariant spacetimes and so massless fields may be described by delta distributions. In the case of Einstein-like gravity problems about non-renormalizability of point-like particles are avoided because of the non-existence of the point-like particles; in the case of conformal gravity renormalizability issues are circumvented because of the conformal invariance. In the Dirac matter field theory, the non-renormalizability that comes from point-like particles is not an issue, and although here other types of non-renormalizability have not been studied, we believe our results to be already interesting.

\end{document}